\documentclass[preprint,proceedings]{rmaa}

\def\fref#1{Figure~\ref{#1}}
\SetYear{2005}
\SetConfTitle{Ninth Texas-Mexico Conference on Astrophysics}

\title{QSO Host Galaxy Luminosity and [OIII] Line Width as a Surrogate
  for Stellar Velocity Distribution} 

\author{E. W. Bonning\altaffilmark{1}, G. A. Shields\altaffilmark{2}, S. Salviander\altaffilmark{2}, R. J. McLure\altaffilmark{3}}

\altaffiltext{1}{ Laboratoire de l'Univers et de ses Th\'{e}ories,
  Observatoire de Paris, F-92195 Meudon Cedex, France,
  (\protect\email{erin.bonnning@obspm.fr}).} 

\altaffiltext{2}{Department of Astronomy, University of Texas at
 Austin, Austin, TX 78712.}

\altaffiltext{3}{Institute for Astronomy, University of Edinburgh,
  Royal Observatory, Edinburgh EH9 3HJ.}

\suppressfulladdresses

\listofauthors{E. W. Bonning, G. A. Shields, S. Salviander, R. J. McLure}
\indexauthor{Bonning, E. W.}
\indexauthor{Shields, G. A.}
\indexauthor{Salviander, S.}
\indexauthor{McLure, R. J.}

\addkeyword{black hole physics}
\addkeyword{galaxies: active}
\addkeyword{quasars: general}

\begin{document}
% Typeset article header
\maketitle 

\boldabstract{Supermassive black holes in galactic nuclei show a close
relationship between the black hole mass
$M_{\mathrm{BH}}$ and the luminosity $L$ and stellar velocity dispersion $\sigma_*$
of the host galaxy bulge.  Probing these relationships at high
redshift may shed light on the link between the formation of the
galactic bulge and central  black hole, but direct
measurements of $\sigma_*$ at 
high redshift are difficult.   We show that  [\ion{O}{iii}] line widths  
provide a useful surrogate for $\sigma_*$ by comparing 
$\sigma_{\mathrm{[O~III]}}$ with the value of $\sigma_*$ predicted by
the Faber-Jackson relation for QSOs with measured host galaxy luminosity.  
Over a wide range of AGN luminosity, $\sigma_{\mathrm{[O~III]}}$ tracks $\sigma_*$,
albeit with considerable scatter.  [\ion{O}{iii}] line widths are narrower by
$~0.1$ dex in radio-loud QSOs than in radio-quiet QSOs of similar  $L_{\mathrm{host}}$. In low redshift QSOs, the ratio of star formation rate to black hole growth rate is  much smaller than the typical ratio of bulge mass to black
hole mass.}   

Nelson \& Whittle (1995, 1996) made a comparison
of bulge magnitudes, [\ion{O}{iii}] line widths, and $\sigma_*$
in Seyfert galaxies, finding on average good agreement between
$\sigma_*$ and $\sigma_{\mathrm{[O~III]}} \equiv \mathrm{FWHM([\ion{O}{iii}])/2.35}$. However, direct comparisons of
$\sigma_{\mathrm{[O~III]}}$  with $\sigma_*$ have generally been limited to
lower luminosity AGN, and it is
important to evaluate the substitution of $\sigma_{\mathrm{[O~III]}} $ for
$\sigma_*$  at higher QSO luminosities. Here we
do this by studying the Faber-Jackson relation \citep{ForbesPonman, Kormendy83}
for a sample of quasars for which host galaxy luminosities are available.

Host galaxy magnitudes for ellipticals, and bulge magnitudes
for spiral hosts were taken from the literature and from our own
unpublished measurements (see Bonning et~al. 2005 for details). The
[\ion{O}{iii}] line width, continuum luminosity, and broad H$\beta$
width were measured  from spectra from
Marziani et~al. (2003) and McLure \& Dunlop (2001).   We assume a
cosmology with $ H_0 = 70~\mathrm{km~s^{-1}~Mpc^{-1}}, \Omega_{\mathrm M} = 0.3,
\Omega_{\Lambda} = 0.7$. 

\begin{figure}[!h]
  \includegraphics[width=\columnwidth]{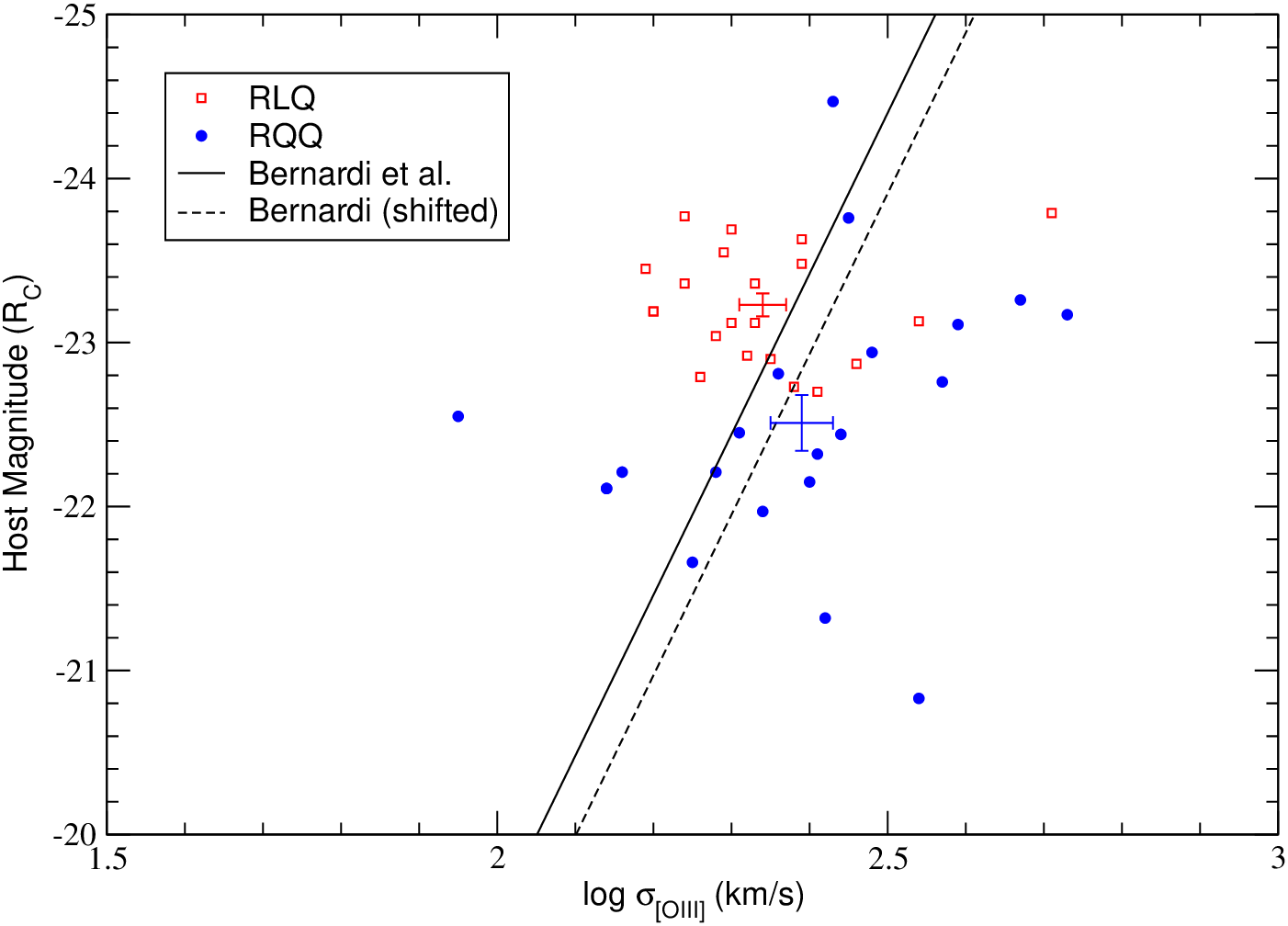}
  \caption{The above plot shows the sample of quasars for  which we
    have host galaxy bulge magnitudes ($R_\mathrm{Cousins}$)
    and  $\sigma_{\mathrm{[O~III]}}$.  The
    straight line is the Faber-Jackson 
relation measured by Bernardi et al. (2003); the dashed line is the 
same relation with log $\sigma$
displaced by 0.05 (see Bonning et al. 2005). The crosses indicate
the mean values and errors of the mean, the RL being above and to the left
of the RQ mean. (From Bonning et al. 2005)}
\label{fig:plot1} 
\end{figure}

\begin{figure}[!t]
  \includegraphics[width=\columnwidth]{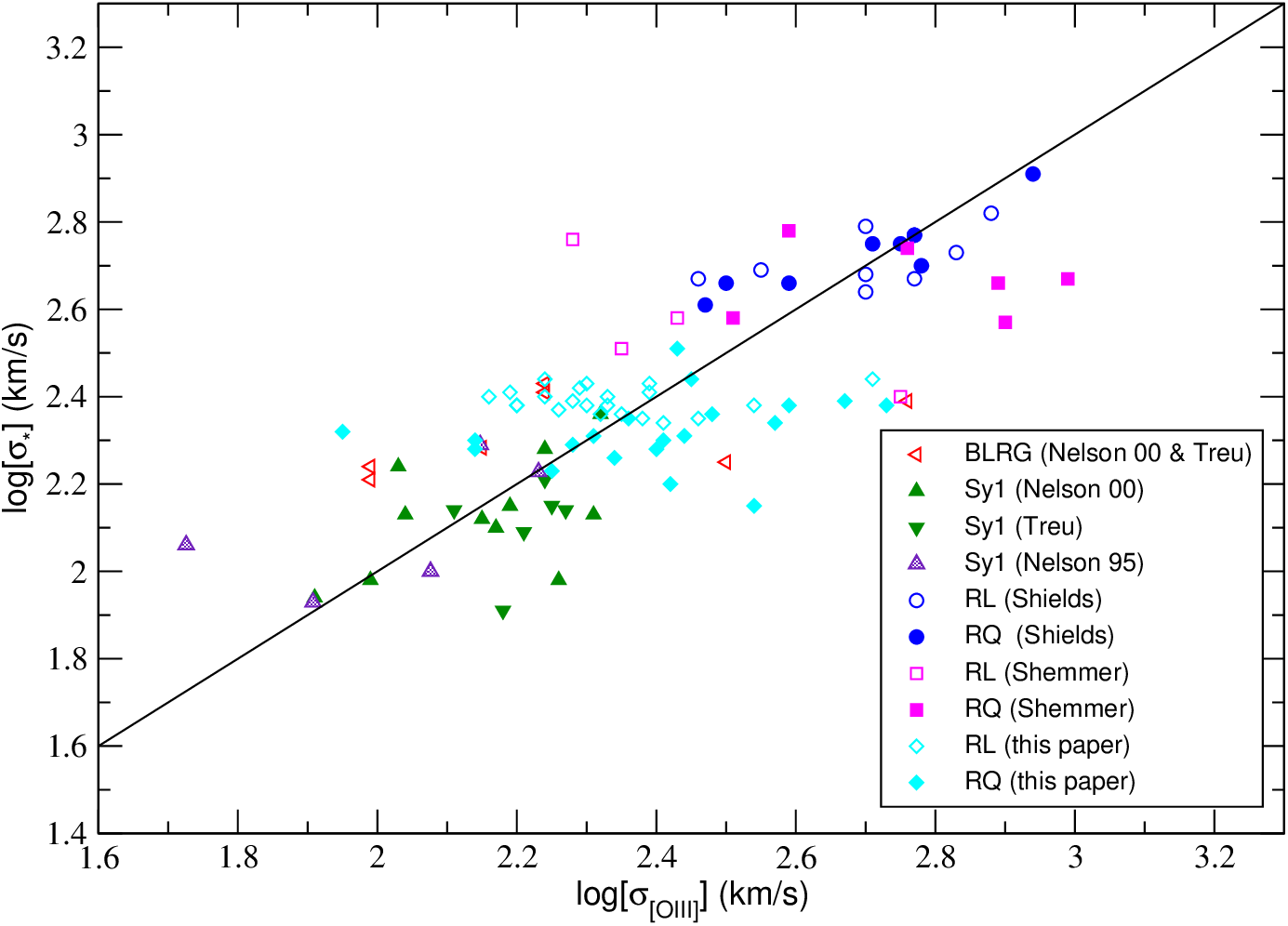}
  \caption{Plot of $\sigma_{\mathrm{[O~III]}}$ v  $\sigma_*$ (measured or inferred)  for our
  data sample and  others.  QSOs from this paper  have  $\sigma_*$ calculated
  from their host  luminosity  via the Faber-Jackson
  relation.  The broad line radio
  galaxies (BLRG) and  Seyferts (see legend) have directly measured
  $\sigma_*$. High luminosity  QSOs from Shields et~al. (2003) and
  Shemmer et~al. (2004) have    $\sigma_*$ calculated from the
  $M_{\mathrm{BH}}$ --  $\sigma_*$ relation 
\citep{Tremaine02}.   (From Bonning et al. 2005)}
 \label{fig:sigsig}
\end{figure}

Our results for host magnitude ($M_{\mathrm{host}}$) and 
$\sigma_{\mathrm{[O~III]}}$, plotted in \fref{fig:plot1}, agree in
the mean with the Faber-Jackson relation. Intrinsic scatter is
$\sim$~0.13 dex in $\sigma_{\mathrm{[O~III]}}$, sufficient to obscure 
the expected increase in $\sigma_{\mathrm{[O~III]}}$ 
over our limited range of $M_{\mathrm{host}}$.  However, 
\fref{fig:sigsig} shows a clear increase in
$\sigma_{\mathrm{[O~III]}}$ with $\sigma_*$ over a much wider range of AGN luminosity,
using objects for which $\sigma_*$ is directly measured or inferred from $L_{\mathrm{host}}$ or $M_{\mathrm{BH}}$. 
  
\vskip20pt

\begin{figure}[!t]
  \includegraphics[width=\columnwidth]{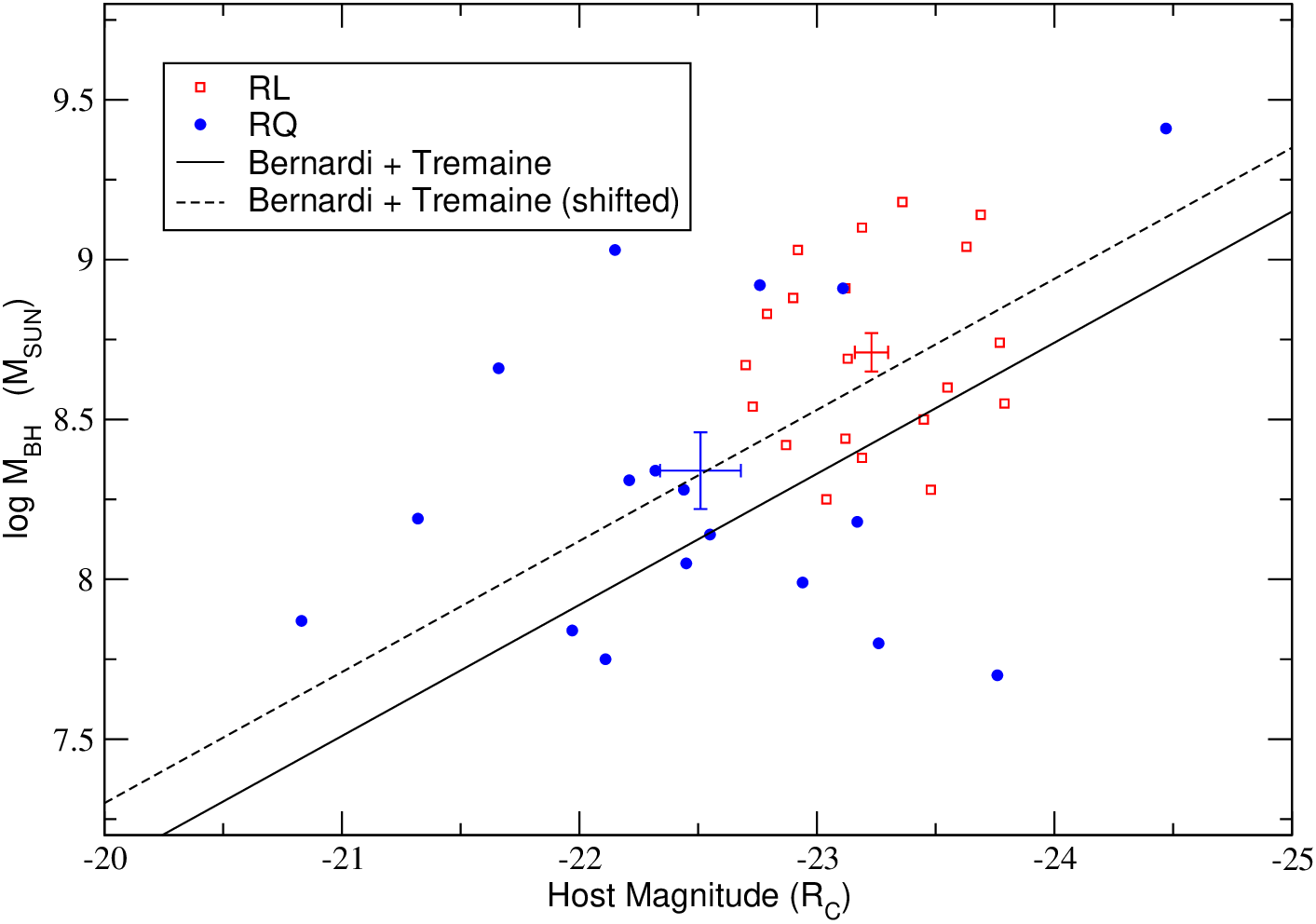}
  \caption{ $M_{\mathrm{BH}}$ versus
    $M_{\mathrm{{host}}}$  for the same objects as \fref{fig:plot1}. It
    can be seen that the RL objects are not offset from the RQ objects
    in relation to the normal $M_{\mathrm{BH}}$~--~$M_{\mathrm{{host}}}$
    trend. (From Bonning et al. 2005) }
 \label{fig:plot3}
\end{figure}

Our RL objects have, on average,
narrower [\ion{O}{iii}] lines than the RQ objects, for a given $L_{\mathrm{host}}$.
A similar RL - RQ
offset has been observed in the $M_{\mathrm{BH}} -
\sigma_{\mathrm{[O~III]}}$  relation for QSOs by Shields et al. (2003) and by
Bian \& Zhao (2004).  The latter
suggested that geometrical effects in RLQ might affect the observed
H$\beta$  widths or continuum luminosity, used to
derive $M_{\mathrm{BH}}$ (Shields et al. 2003). 
However, a comparison of
$M_{\mathrm{BH}}$ with $M_{\mathrm{{host}}}$  (\fref{fig:plot3}) shows no
significant offset of RL objects relative to the expected slope. This
suggests that narrower  $\sigma_{\mathrm{[O~III]}}$  for RL objects is
responsible for the RL-RQ offset, and not any systematic effect
involving $M_{\mathrm{BH}}$. 

The proportionality of black hole mass and bulge mass,
$M_{\mathrm{BH}} \approx 0.0013 M_{\mathrm{bulge}}$ (Kormendy \& Gebhardt 2001) raises
the question of whether black hole growth and bulge growth occur
simultaneously. The average bolometric luminosity of
our RQQ ($L \approx 10^{45.7}~\mathrm{erg~s^{-1}}$)  corresponds to an
accretion rate $\dot M \approx 1~M_\odot\,\mathrm{yr ^{-1}}$ for an
efficiency $L \approx 0.1\dot M c^2$.  The corresponding star formation rate 
is $\sim 700~M_\odot~\mathrm{yr^{-1}}$ to maintain the black hole --
bulge relationship.  Such rates are observed in some 
ULIRGs but not in the PG QSOs of our sample.  Ho (2005)
finds that star formation  is supressed in PG QSOs,
despite abundant molecular gas.
Star formation in PG QSOs is far less than required to
maintain detailed balance between bulge and black hole mass.  However,
the main growth of the black hole at higher redshifts may involve more
nearly simultaneous star formation and bulge growth.

\acknowledgements
EWB was supported by a NASA GSRP fellowship and a Chateaubriand fellowship. 
GAS and SS were supported
by Texas Advanced Research Program grant 003658-0177-2001 and NSF grant
AST-0098594.  RJM acknowledges the support of the Royal Society.

\end{document}